\documentclass[10pt,letterpaper]{article}
\usepackage[top=0.85in,left=2.75in,footskip=0.75in]{geometry}

\usepackage{amsmath,amssymb}

\usepackage{changepage}

\usepackage{textcomp,marvosym}

\usepackage{cite}

\usepackage{nameref,hyperref}

\usepackage[nopatch=eqnum]{microtype}
\DisableLigatures[f]{encoding = *, family = * }

\usepackage[table]{xcolor}

\usepackage{array}

\usepackage{subcaption}
\usepackage{graphicx}
\usepackage{xcolor}

\newcolumntype{+}{!{\vrule width 2pt}}

\newlength\savedwidth

\raggedright
\usepackage[aboveskip=1pt,labelfont=bf,labelsep=period,justification=raggedright,singlelinecheck=off]{caption}

\makeatletter
\renewcommand{\@biblabel}[1]{\quad#1.}
\makeatother

\usepackage{lastpage,fancyhdr,graphicx}
\usepackage{epstopdf}
\fancyheadoffset[L]{2.25in}
\fancyfootoffset[L]{2.25in}
\begin{document}
\vspace*{0.2in}

\begin{flushleft}
{\Large
\textbf\newline{An affinity based opinion dynamics model 
for the evolving pattern of political polarization} 
}
\newline
\\
Xiaoming J. Zhang\textsuperscript{1,2*},
Yuzhong Hu\textsuperscript{1}, and
Yiming Zhang\textsuperscript{1}

\bigskip
\textbf{1} Beijing Institute of Mathematical Sciences and Applications, Beijing, China
\\
\textbf{2} Institute for Applied Mathematics, Tsinghua University, Beijing, China
\\
\bigskip
*Corresponding author: Xiaoming J. Zhang, zhangxiaoming@bimsa.cn
\end{flushleft}
\section*{Abstract}
Political polarization has been a subject that has attracted many studies in recent years. We have developed an opinion dynamics model with affective homophily effect and national social norm effect to describe this phenomenon. The time evolution of the polarization between the two parties and the spread of opinions within each party are affected by three factors: the repulsive effect between the two parties, the attractive and repulsive effects between the members in each party, and the national social norm effect that pulls the opinions of all members towards a common norm. The model is internally consistent and is applied to the simulation of the symmetric patterns of polarization and spread of the opinion distributions in the U.S. Congress and the results align well with 154 years of recorded data. The time evolution of the strength of the national social norm effect is obtained and is consistent with the important historical events occurred during the past one and half century.


\section{Introduction}
The political landscape of the United States has become increasingly polarized over the past four decades \cite{McCarty2016}. Researchers have suggested that while the polarization is undermining democracy and legislative effectiveness, understanding the phenomenon could ultimately reveal strategies for bridging the divides \cite{Axelrod2021,Hill2015,Balietti2021}. However, despite past efforts in studying political polarization, most analytical models are fundamentally limited in capturing the interactions among congress members and the historical context. It is therefore appealing for us to raise the following questions: What drives the separation and reunion of ideologies in the U.S. political system? Is there a relationship between partisan polarization and critical historical events?

Political polarization arises as the spectrum of public opinion fractures or as differences in perspectives sharpen. Jost et al. \cite{Jost2022} has delineated that affective polarization is characterized by the strong emotional responses, whether positive or negative, that members of different social groups evoke in each other. As an important psychological mechanism that plays a pivotal role in driving polarization, group justification encompasses the collective inclination to promote the benefit of one's own group while opposing rival groups. Cole et al. \cite{Cole2023} have reviewed the works on political polarization related to climate change issues. They emphasize two types of mechanisms of political polarization: individual-level psychological processes and group-level psychological processes. The former drives the polarization through ideology, personality traits and cognitive styles, and perception of risks, threats, and morality. The latter includes social identities, social norms, and affective polarization.

Researchers have been attempting to explain political polarization with agent-based models and data simulations. These models usually adopt a utility maximization approach which assumes that the agents, affected by various types of public influences, selfishly make decisions to maximize their utility. Depending on the problem of interest, these models study the temporal evolution of opinion distribution among all Congress members, either in a discrete or continuous form, by either a deterministic or stochastic process. One common assumption is that the political parties make decisions to maximize their vote counts. Downs \cite{Downs1957} has modeled the competition of two parties for voters, where the voters' opinions follow an invariant zero-mean Gaussian distribution, and each voter votes for a party to maximize an expected utility. Each party adjusts its opinion position to maximize the expected number of votes it receives. The Downsian model predicts that the two parties should reach a consensus at the median opinion position of all voters. However, this conclusion does not coincide with the reality, as partisan polarization in the U.S. has been a well observed phenomenon.

A notable variant of the Downsian model is the satisficing model developed by Yang et al. \cite{Yang2020}. The time evolution of a party’s opinion can be described using two characteristic variables: the party’s opinion as the average opinion of its members and the party’s opinion spread as the standard deviation of its members. Like the Downsian model, the opinion of a party moves in a direction that maximizes the expected number of votes. At the same time, a voter decides which party to vote for by randomly selecting a satisficing party (or abstaining from voting if there is none). Through data validation, the satisficing model captures opinion polarization between the U.S. Democratic and Republican parties since 1961. The model also develops a relationship between partisan polarization and opinion spread and predicts that the opinions of the members of each party are more centralized as the two parties become more polarized, which agrees well with the observed data. However, the work does not provide an explanation on how the polarization is developed and why it exhibits a wavy ideology distribution pattern over the past one and a half century of the U.S. Congress.

Recently, Lanzetti et al. \cite{Lanzetti2022} have extended the satisficing model to predict the complete opinion distribution of a party using Wasserstein gradient flows. The model predicts that the opinion distributions of the two parties should become more polarized and homogeneous within each party with time, converging to asymmetric distributions. However, while the extended satisficing model captures the overall tendency of partisan polarization, it does not explain notable exceptions where the opinion distributions change with time in a wavy fashion over the long history of the U.S. Congress. The author indicates that these exceptions are due to impact factors such as historical context, election rounds, and political campaigns without elaborated analysis.

Jones et al. \cite{Jones2022} have introduced a new way to define voters’ distribution and utilities, where the voters’ population is composed of two subpopulations with polarized centers. Partisan polarization would exceed subpopulation polarization either when the two subpopulations are homogeneous and polarized or when they are heterogeneous and centralized. Ferri et al. \cite{Ferri2022} have introduced a three-state model that borrows ideas from thermodynamics to opinion dynamics. The system is analogized with a thermal bath of a specific temperature, representing social agitation that affects the stochastically evolving dynamics of the system. The result shows that the system converges to a disordered state with polarized opinion clusters when the temperature is high and the neutrality parameter is small, or to a relatively unified state when the opposite is true. This corresponds closely to the relationship between population spread and partisan polarization in \cite{Jones2022}.

Opinion dynamics models have been used in studying affective polarization. The interactions of people with close opinions would lead to agreement and positive affection, while interactions of people with distant opinions result in distrust and negative affection \cite{Balietti2021,Liu2015}. Iyengar et al. \cite{Iyengar2019} have traced affective polarization to the power of partisanship as a social identity. Finkel et al. \cite{Finkel2020} have utilized a ``feeling thermometer'' to measure the out-party hate level and find that it is the strongest in America than in eight other nations. Lu et al. \cite{Lu2019} have studied a dynamic system of the conversion between polarization and cooperation in political interactions. The system has used data from roll-call votes cast in the U.S. Congress and showed a growth of polarization over the recent decades. However, the model itself does not explain the cause of the wavy political polarization pattern in a longer time scale.

Leonard et al. \cite{Leonard2021} have applied a nonlinear opinion dynamics model to study partisan asymmetry and polarization. In the model, the two parties adjust their opinion positions based on self-reinforcement response mechanisms, in which a party exacerbates its polarized position to gain support. The model is tested using opinion data of the U.S. Congress since 1959 by searching for optimal parameters. The paper is significant because it offers a plausible explanation of the two parties’ polarization asymmetry based on the changes in the public’s opinion. Moreover, the model is robust and does not rely on fine-tuning specific parameters to capture the overall tendency of opinion shifts. However, the authors do not attempt to validate their model with data before 1951, which exhibits more diverse behaviors of opinion shifts. In fact, since the self-reinforcement mechanism always results in opinion polarization, the model cannot capture other types of behaviors, such as the centralization of two parties’ opinions around World War II.

Baldassarri and Bearman \cite{Baldassarri2007} have studied the paradox of the simultaneous absence and presence of attitude polarization and the paradox of the simultaneous presence and absence of social polarization. Later, Baldassarri and Page \cite{Baldassarri2021} have reviewed this work in light of the theoretical distinction between ideological partisanship, which is generally rooted in sociodemographic and political cleavages, and affective partisanship, which is, instead, fueled mainly by emotional attachment and repulsion, rather than ideology and material interests.

This work intends to study the evolving polarization of the parties and their spread with the effects of in-party and cross-party opinion dynamics, and the effect of a time-dependent national social norm. Based on a micro-scale agent-based model, we derive an analytical theory on how the two parties’ opinions influence each other with the presence of the national social norm effect. Four parameters are used in the theory: (1) a tolerance opinion difference parameter that determines whether the mutual impact of any two interacting individuals is attractive or repulsive; (2) an influence decay parameter whose inverse determines the exponential decay rate of their mutual impact as their opinion difference increases; (3) a pre-coefficient and a rate for the exponential increase of the opinion exchange efficiency due to the fast progress of communication technology in the past; and (4) a time-dependent function characterizing the strength of the national social norm effect. These four parameters and the strength function are obtained by fitting the theory with empirical data from a frequently used dataset \cite{Lewis2021}. The obtained values of these four parameters and their social physical meanings can be reasonably well interpreted. The strength of the national social norm effect is consistent with the major historical events occurred in the past one and a half century.

This paper is organized as follows. Section 2 describes the ideology dataset we used for the U.S. Congress \cite{Lewis2021}. Section 3 develops the theory for the evolution of polarization and spread with proper assumptions and approximations. Section 4 describes the numerical method used to assimilate theoretical model and observational data. Section 5 presents the analysis results and interpretations, and the justification of the approximations used. Section 6 summarizes the contributions of this work and points out the areas for further research.

\section{Data description}

We apply the ideology dataset from the U.S. Congress \cite{Lewis2021} for this research. The dataset comprises two-dimensional ideology scores of congressional members from 1868 to 2022 computed by the Dynamic, Weighted, Nominal Three-Step Estimation (DW-NOMINATE) algorithm \cite{Boche2018}. We choose the scores in the first dimension (economic liberalism-conservatism) of this dataset to represent the members’ opinion distributions, which has data every two years with 78 time points, or Congress sessions, covering 154 years.

\begin{figure}[ht]
\centering
\includegraphics[width=\textwidth]{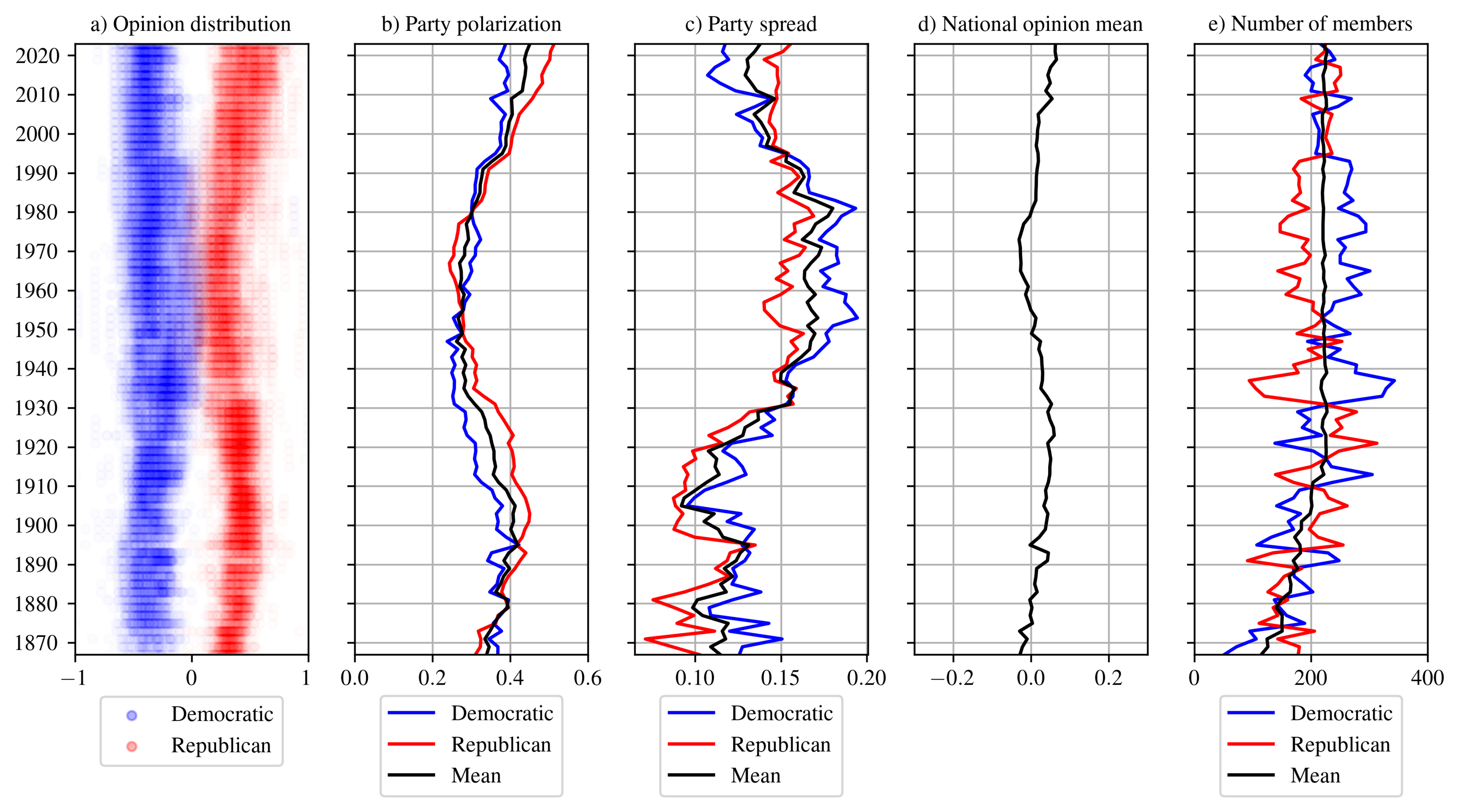}
\caption{(a) Opinion distributions of the Democratic and Republican party members from 1868 to 2022; (b) Absolute values of the polarizations of the two parties and their mean; (c) Spreads of the two parties and their mean; (d) The average of the opinion means of the two parties; (e) The numbers of congressmen in the two parties and their means.}
\label{fig1}
\end{figure}

Fig.~\ref{fig1}(a) shows the opinion distribution of the Democratic (blue) and Republican (red) parties representing a wavy pattern of separation and unification process. The bipartisan polarizations can be defined as the means of the opinion distributions of the two parties. The absolute values of the polarizations are as shown in Fig.~\ref{fig1}(b), which are stronger around 1880 to 1910 and 2000 to 2020, and weaker around 1930 to 1980. The opinion spreads in Fig.~\ref{fig1}(c) of the two parties, computed as the standard deviation of the opinion distributions about their corresponding means, are narrower when there is an intense polarization and are wider when the polarization is weak.

It is worth noting the significant asymmetric disparities in both the polarizations and spreads of the two parties. In Fig.~\ref{fig1}(d), the average of the two parties’ opinion means slightly deviates from the zero-opinion level. Additionally, in Fig.~\ref{fig1}(e), the numbers of congress members of the two parties fluctuate over time, with their mean increasing from 1868 to 1920 and stabilizing thereafter. This work focuses on modeling the symmetric polarization and spread patterns of the two parties, as well as their long-term temporal wavy evolutions, leaving the study of asymmetry patterns to future research.

\section{Theory}

Let $D$ represent the Democratic party and $R$ represent the Republican party in the U.S. Congress. An individual member $i$'s opinion level at time $t$, denoted as $B_i(t)$, resides within a one-dimensional segment from $-1$ to $+1$. The opinion of each member at a given time is affected by opinion impacts from all other individuals and a national social norm effect. The general governing equation for the evolution of $B_i(t)$ is
\begin{equation}
\frac{dB_i(t)}{dt} = \sum_{j \in D} I_{ji} + \sum_{j \in R} I_{ji} - \alpha(t)[B_i(t) - B_o(t)].
\label{eq:governing}
\end{equation}

In Eq.~\eqref{eq:governing}, the first term on the right-hand side is the summation of impacts from all individuals in the Democratic party ($D$), and the second term is from all individuals in the Republican party ($R$). The opinion impact $I_{ji}$ of an individual $j$ on $i$ is defined by
\begin{equation}
I_{ji} = A(t) D_{ji} e^{-\frac{|B_j - B_i|}{B_H}} (B_j - B_i).
\label{eq:impact}
\end{equation}

The opinion impact is a product of the opinion difference between the two individuals and three additional factors. The factor $D_{ji} e^{-|B_j - B_i|/B_H}$ represents affinity, a measure of likeness between two individuals, which decreases as the opinion difference between them increases. $D_{ji}$ is expressed using an opinion influence tolerance parameter, $B_T$:
\begin{equation}
D_{ji} = 
\begin{cases}
1 - \frac{|B_j - B_i|}{B_T}, & \text{if } |B_j - B_i| < 2B_T, \\\\
-1, & \text{if } |B_j - B_i| \geq 2B_T.
\end{cases}
\label{eq:tolerance}
\end{equation}
The tolerance parameter determines the threshold of the opinion difference between two individuals from exerting attractive or positive influence to repulsive or negative influence.

The factor $e^{-\frac{|B_j - B_i|}{B_H}}$ represents the homophily effect quantifying the extent of mutual influence between two individuals, which is also related to the opinion difference. $B_H$ is an influence decay parameter; its inverse determines the rate of exponential influence decay as the opinion difference increases. When the opinion difference between two individuals becomes large compared to $B_H$, the mutual influence power between them becomes small. This decay parameter characterizes the range of individuals in a population that may have a significant influence on an individual, whether positive or negative. In the case of positive influence, this homophily effect has been validated by He and Zhang \cite{He2023} using experimental data. In this work, we assume that each of $B_T$ and $B_H$ is the same for both in-party and cross-party impacts.

The proportional factor $A(t)$ relates to the interaction efficiency between individuals within the opinion dynamic system. The growth of this efficiency stems from advancements in transport technologies (such as automobiles, highways, aircrafts) and communication technologies (such as radios, TVs, phones, cell phones, and the Internet) over the past one and a half century, resulting in an exponential growth.

The third term in Eq.~\eqref{eq:governing} characterizes the national social norm effect, affecting all individuals across both parties, a topic extensively studied in political science literatures \cite{Cole2023}. In Eq.~\eqref{eq:governing}, $B_o$ denotes the position of the norm, and $\alpha(t) > 0$ represents the strength of the norm effect. This strength is related to important world events occurred around or before time $t$ and is therefore time-dependent. It is strong in the presence of a common national threat for both parties and weak when such threats are absent.

The model Eqs.~\eqref{eq:governing}--\eqref{eq:tolerance} are extensions of the widely used classical DeGroot model \cite{DeGroot1974} in opinion dynamics studies. The extensions are made in three aspects: (1) In addition to the opinion difference between two individuals, the impact of one person on another also depends on their mutual likeness (affinity); (2) The mutual influence of individuals' opinions also depends on the interaction efficiency, which is assumed to be increasing exponentially over the past one and a half century; (3) Changes of people's opinions are also regulated by a national social norm effect, which tends to strengthen when the nation faces a common threat and weaken during peaceful times.

The members of the U.S. Congress change from session to session, with each session lasting two years. It is difficult to identify a consistent counterpart across sessions for each member with similar opinion position. This discontinuity prevents us from tracking their individual opinion trajectories over time. To overcome this limitation, we adopt a mean-field-like approach and study the evolutions of key quantities of the opinion distributions in the following analysis.

In this work, the term ``partisan polarization'' (or ``polarization'') refers to the deviation of each party’s opinion center or mean opinion level from the national norm, while ``opinion spread'' (or ``spread'') indicates the standard deviation of opinion distribution within each party, also termed inclusiveness in some literature \cite{Yang2020,Lanzetti2022}. Using shorthand notations $B_i$ for $B_i(t)$, the partisan polarizations of the Democratic ($D$) and Republican ($R$) parties, respectively, are expressed as
$B_D = \frac{1}{N_D} \sum_{i \in D} B_i < 0$, 
$B_R = \frac{1}{N_R} \sum_{i \in R} B_i > 0.$
Their time-varying absolute values and average are shown in Fig.~\ref{fig1}(b) above. The squares of the opinion spreads in the two parties are represented by
$\sigma_D^2 = \frac{1}{N_D} \sum_{i \in D} (B_i - B_D)^2,$ $\sigma_R^2 = \frac{1}{N_R} \sum_{i \in R} (B_i - B_R)^2,$
where $N_D$ and $N_R$ are the numbers of congressmen in the Democratic party $D$ and Republican party $R$, respectively. The time-varying opinion spreads of the two parties and their average are shown in Fig.~\ref{fig1}(c).

We focus on the temporal evolution of polarization and spread of the two parties instead of tracking the opinion evolution of the individual congress members. Noting that the total number of congress members remains fixed after 1920, while the number of each party fluctuates. These fluctuations are modest compared to the total number, apart from exceptional periods such as World War II, as shown in Fig.~\ref{fig1}(e). Since the overall opinion impact on any member is the summation of the impacts from all other members, these fluctuations have limited impact on the qualitative understanding of the mean-field opinion dynamics. Therefore, to reduce the complexity associated with short timescale phenomena, we assume that the numbers of congressmen in the two parties are the same for all years, namely $N_D = N_R = N$.

We also assume that the opinion position of the national social norm is fixed at $B_o(t) = 0$, and that the opinion distributions of the two parties are symmetric with respect to this national norm. The symmetry assumption is supported by the empirical observation that, over long timescales, the polarization and spread of the two parties are generally symmetric, as illustrated in Fig.~\ref{fig1}(b) and Fig.~\ref{fig1}(c). This assumption has also been widely used in prior studies \cite{Downs1957,Yang2020,Lanzetti2022,Jones2022,Ferri2022,Liu2015,Iyengar2019,Finkel2020,Lu2019}, given the noisy nature of social science datasets. Moreover, this assumption enables the development of a tractable model capable of capturing the dominant, long-term aggregated trends over the 154-year study period. Denoting $\mu$ and $\sigma$ as the polarization and the spread of opinion distribution, we have
$
B_R = -B_D = \mu > 0, \quad \sigma_R^2 = \sigma_D^2 = \sigma^2.
\label{eq:symmetry}
$

We further make the following four approximations, which will be justified in Section~5. It is noted that these approximations are not perfectly accurate for all in-party and cross-party interactions due to the existence of members with extreme opinions. Through these approximations, we aim to achieve a model that captures the variation of polarization and spread over the long term using simple governing equations.

\textbf{Approximation 1}: The in-party exponential decaying effect is negligible since $|B_j - B_i| \sim \sigma < B_H/3$ for two individuals with the same party, $i, j \in D$ or $i, j \in R$, so that
$e^{-\frac{|B_j - B_i|}{B_H}} \approx 1.$
This is a simplification that holds when most in-party opinion differences lie within a narrow range compared to $B_H$. We acknowledge that this assumption may not hold during several short periods with extreme party opinion spread. For the purpose of a qualitative understanding of the general long-term trend, the effect of such extreme cases is neglected.

\textbf{Approximation 2}: For any two individuals $i$ and $j$ from the same party, we assume $|B_j - B_i| < 2B_T$. Therefore,
$
I_{ji} = A(t) \left( 1 - \frac{|B_j - B_i|}{B_T} \right)(B_j - B_i).
$
This approximation allows that members of the same party with close opinions ($|B_j - B_i| < B_T$) interact attractively, while those with opinion difference $B_T < |B_j - B_i| < 2B_T$ interact repulsively. The multiplication factor $D_{ji}$ in Eq.~\eqref{eq:tolerance} is thus always greater than $-1$.

\textbf{Approximation 3}: The cross-party exponential decaying effect is important, since $|B_j - B_i| \approx B_R - B_D = 2\mu$ for two individuals $i$ and $j$ belonging to different parties, and $2\mu > 2B_T$. Thus,
\begin{equation}
I_{ji} = I_{RD} = A(t) \cdot (-1) \cdot e^{-\frac{|B_R - B_D|}{B_H}} (B_R - B_D) = -2A(t)\mu e^{-2\mu/B_H}.
\label{eq:crossparty}
\end{equation}

\textbf{Approximation 4}: The proportional factor $A(t)$ is assumed to be an exponential function of time $t$ to describe the improvement of interaction efficiency due to the advancement of transportation and communication technologies. This aligns with recent findings on how digital media could speed up the frequency of communications and intensify political polarization \cite{Balietti2021,Gentzkow2016}.

Adopting these approximations, for an individual $i$ in the $D$ party with $B_i < 0$, Eq.~\eqref{eq:governing} becomes
\begin{equation}
\frac{dB_i}{dt} = A(t) \sum_{j \in D} \left(1 - \frac{|B_j - B_i|}{B_T}\right)(B_j - B_i) - 2A(t)N\mu e^{-2\mu/B_H} - \alpha(t) B_i.
\label{eq:indivD}
\end{equation}

The rate equation for partisan polarization $\mu = -\frac{1}{N}\sum_{i \in D} B_i$ can be obtained by averaging Eq.~\eqref{eq:indivD} over all individuals in the $D$ party, recognizing that the first term vanishes after summation:
\begin{equation}
\frac{d\mu}{dt} = \mu \left[2A(t)N e^{-2\mu/B_H} - \alpha(t)\right].
\label{eq:polarization_evolve}
\end{equation}

The opinion centers of the two parties converge or diverge depending on the competing repulsive effect between the two parties and the attraction effect of the national norm. Specifically, the bipartisan polarization increases when the national norm strength $\alpha(t)$ is smaller than the threshold $2A(t)N e^{-2\mu/B_H}$, and vice versa.

The equation for the spread of the parties, described by the standard deviation of the opinion distribution within each party, can be obtained as shown below, with derivations detailed in Appendix~A:
\begin{equation}
\frac{d\sigma}{dt} = \sigma \left[A(t)N\left(\frac{4\sigma}{\sqrt{\pi} B_T} - 1\right) - \alpha(t)\right].
\label{eq:spread_evolve}
\end{equation}

The evolution of the spread of party opinion is controlled by two effects. The first term represents the divisive effect within the party when $\frac{4\sigma}{\sqrt{\pi} B_T} > 1$. The second term is the national social norm effect, which tends to reduce the spread. The spread increases when the national norm strength $\alpha(t)$ is smaller than $A(t)N\left(\frac{4\sigma}{\sqrt{\pi} B_T} - 1\right)$, and decreases otherwise.

The mean-field-like approach, with Eq.~\eqref{eq:polarization_evolve} and Eq.~\eqref{eq:spread_evolve} as the two governing equations, provides a mechanistic understanding of how the long-term evolutions of polarization and spread are shaped by social psychological processes, the effect of the national social norm—which is expected to be related to important historical events—and the enhancement of interaction efficiency.

To mitigate strong polarization and spread, three strategies may be considered based on the governing equations Eq.~\eqref{eq:polarization_evolve} and Eq.~\eqref{eq:spread_evolve}:
(1) Fostering a shared civic identity and emphasizing widely endorsed societal values among all citizens, as supported by findings in political psychology and sociology \cite{VanDerLinden2016}. This corresponds to the increase of the strength of the national social norm effect $\alpha(t)$ in both equations.
(2) Promoting respectful discourse and exposure to ideologically diverse individuals across the two parties, which may decrease mutual hostility \cite{Axelrod2021,Balietti2021}. This corresponds to an increase of the homophily decay parameter $B_H$ in Eq.~\eqref{eq:polarization_evolve}, for an increase of affinity.
(3) Encouraging interactions within each party among those with different views, which is expected to decrease the spread by fostering increased tolerance \cite{Axelrod2021,Balietti2021}. This corresponds to an increase of the tolerance parameter $B_T$ in Eq.~\eqref{eq:spread_evolve}.

\section{Numerical method for the assimilation of data and theory}

In their discrete forms using a forward difference scheme, Eq.~\eqref{eq:polarization_evolve} and Eq.~\eqref{eq:spread_evolve} can be written as
\begin{equation}
\mu(t+\Delta t) = \mu(t) + \mu(t)\left[2\bar{A}(t)e^{-\frac{2\mu(t)}{B_H}} - \alpha(t)\right]\Delta t,
\label{eq:mu_discrete}
\end{equation}
and
\begin{equation}
\sigma(t+\Delta t) = \sigma(t) + \sigma(t)\left[\bar{A}(t)\left(\frac{4\sigma(t)}{\sqrt{\pi} B_T} - 1\right) - \alpha(t)\right]\Delta t,
\label{eq:sigma_discrete}
\end{equation}
where $\bar{A}(t) = A(t)N\Delta t$ represents the total amount of influence from all individuals impacting on the system, and $\Delta t$ is the incremental time step, which is two years in this study.

The time variable $t$ starts from the year 1868 and ends in the year 2022. We denote $t_k = 1868 + (k-1)\Delta t$, $1 \leq k \leq K$, as the $k$-th time point, where $K=78$ is the total number of time points. 
We assume $\bar{A}(t_k) = A_0 e^{ck/K},$
with two constants $A_0 > 0$ and $c > 0$, whose values are to be determined after fitting the theory with data.

We aim to optimally utilize information from both the empirical data and the theoretical relationships among the variables and the parameters provided by Eq.~\eqref{eq:mu_discrete} and Eq.~\eqref{eq:sigma_discrete}. The model equations involve four constant parameters $A_0$, $c$, $B_T$, and $B_H$, and an unknown time-dependent national norm strength function $\alpha(t)\Delta t$. By fitting the data with the model using the assimilation algorithm described below, we can obtain the estimates of the time-dependent patterns of polarization and spread, the values of the four parameters, and the intensity function of the national social norm.
For each Congress session $t$, denote $\tilde{\mu}(t)$ and $\mu^{*}(t)$ as the estimated and observed partisan polarization, respectively, and $\tilde{\sigma}(t)$ and $\sigma^{*}(t)$ as the estimated and observed spread, respectively.

The raw observed data $\mu^{*}(t)$ and $\sigma^{*}(t)$ are very noisy containing short-term fluctuations from session to session over the long duration of 154 years. It is crucial to filter out these short-term noises with the support of the two long-term dynamics equations for the accurate capturing of the long-term trends in polarization and spread. The estimation procedure below, which simultaneously minimizes four loss terms, two for the matching of the data and two for the satisfaction of the equations, allows us to assimilate long-term information from raw observation data with the constraints of the long-term model equations. This strategy is frequently used in data-model assimilation literature and inverse modeling literature, for examples, simultaneous solution of inverse problem governed by PDEs and state estimation~\cite{Flavien2025StateParamEst}, physics-informed neural networks for solving differential equations with unknown parameters using observational data~\cite{Raissi2019PINN}, and variational inference using optimization for the joint estimation of system state and noise parameters~\cite{Lan2024NoiseIden}. 

We seek to reduce the mean squared errors in polarization and spread, in terms of both the data difference between the estimated and the observed, and the mean squared residual errors of the two governing equations, using a minimization procedure. The four mean squared errors are defined as:
We seek to reduce the mean squared errors in polarization and spread, in terms of both the data difference between the estimated and the observed, and the mean squared residual errors of the two governing equations, using a minimization procedure. The four mean squared errors are defined as
\begin{subequations}\label{eq:mse}
\begin{align}
&\mathrm{MSE}_{\mathrm{data}}^{\mu} = \frac{1}{K} \sum_{k=1}^K \left[\tilde{\mu}(t_k) - \mu^{*}(t_k)\right]^2, \label{eq:mse_a} \\
&\mathrm{MSE}_{\mathrm{data}}^{\sigma} = \frac{1}{K} \sum_{k=1}^K \left[\tilde{\sigma}(t_k) - \sigma^{*}(t_k)\right]^2, \label{eq:mse_b} \\
&\mathrm{MSE}_{\mathrm{eq}}^{\mu} = \frac{1}{K-1} \sum_{k=1}^{K-1} \left[\tilde{\mu}(t_{k+1}) - \tilde{\mu}(t_k) - \tilde{\mu}(t_k)\left(2\bar{A}(t_k)e^{-2\tilde{\mu}(t_k)/B_H} - \alpha(t_k)\Delta t\right)\right]^2, \label{eq:mse_c} \\
&\mathrm{MSE}_{\mathrm{eq}}^{\sigma} = \frac{1}{K-1} \sum_{k=1}^{K-1} \left[\tilde{\sigma}(t_{k+1}) - \tilde{\sigma}(t_k) - \tilde{\sigma}(t_k)\left(\bar{A}(t_k)\left(\frac{4\tilde{\sigma}(t_k)}{\sqrt{\pi}B_T} - 1\right) - \alpha(t_k)\Delta t\right)\right]^2. \label{eq:mse_d}
\end{align}
\end{subequations}

The unknown time-dependent national norm strength function $\alpha(t)\Delta t$ is approximated using a linear expansion in terms of $n$ Legendre polynomials with coefficients $(a_0, a_1, a_2, \dots, a_n)$. Introducing a transformation $x_k = \frac{2k}{K} - 1 \in [-1, 1]$ and defining $\tilde{\alpha}(x_k) = \alpha(t_k)\Delta t$, we have the expansion
$
\tilde{\alpha}(x_k) = \sum_{m=0}^{n} a_m P_m(x_k),
$
where $P_m(x)$ is the Legendre polynomial of degree $m$. Thus, the set of parameters to be estimated can be denoted as
$
\Theta = \{A_0, c, B_T, B_H, a_0, a_1, \dots, a_n\}.
$
The seemingly large number of parameters is mainly due to the need to describe the national norm strength function, which is not known a priori and cannot currently be captured by a small set of parameters.

The mean polarization and mean spread across all time points are respectively
\begin{align*}
  \bar{\mu} = \frac{1}{K} \sum_{k=1}^K \mu^{*}(t_k) \approx 0.347, \\
\bar{\sigma} = \frac{1}{K} \sum_{k=1}^K \sigma^{*}(t_k) \approx 0.139.  
\end{align*}
The mean variations per time step for polarization and spread are respectively
\begin{align*}
\bar{\mu}_d = \frac{1}{K-1} \sum_{k=1}^{K-1} |\mu^{*}(t_{k+1}) - \mu^{*}(t_k)| \approx 8.96 \times 10^{-3}, \\
\bar{\sigma}_d = \frac{1}{K-1} \sum_{k=1}^{K-1} |\sigma^{*}(t_{k+1}) - \sigma^{*}(t_k)| \approx 5.36 \times 10^{-3}.
\end{align*}

Since the four mean squared error terms in Eq.~\eqref{eq:mse_a}--\eqref{eq:mse_d} have different orders of magnitude, we use the following normalization factors when constructing a total loss function for minimization:
\[
C_{\mathrm{data}}^{\mu} = \frac{1}{K} \sum_{k=1}^K \left[\mu^{*}(t_k) - \bar{\mu}\right]^2 = 3.03 \times 10^{-3},
\quad \text{for} \quad \mathrm{MSE}_{\mathrm{data}}^{\mu},
\]
\[
C_{\mathrm{data}}^{\sigma} = \frac{1}{K} \sum_{k=1}^K \left[\sigma^{*}(t_k) - \bar{\sigma}\right]^2 = 5.81 \times 10^{-4},
\quad \text{for} \quad \mathrm{MSE}_{\mathrm{data}}^{\sigma},
\]
\[
C_{\mathrm{eq}}^{\mu} = \frac{1}{K-1} \sum_{k=1}^{K-1} \left[\left|\mu^{*}(t_{k+1}) - \mu^{*}(t_k)\right| - \bar{\mu}_d\right]^2 = 4.61 \times 10^{-5},
\quad \text{for} \quad \mathrm{MSE}_{\mathrm{eq}}^{\mu},
\]
\[
C_{\mathrm{eq}}^{\sigma} = \frac{1}{K-1} \sum_{k=1}^{K-1} \left[\left|\sigma^{*}(t_{k+1}) - \sigma^{*}(t_k)\right| - \bar{\sigma}_d\right]^2 = 1.77 \times 10^{-6},
\quad \text{for} \quad \mathrm{MSE}_{\mathrm{eq}}^{\sigma}.
\]

The total loss function is defined as a linear combination of the four mean squared errors described in Eq.~\eqref{eq:mse}, and is a function of the polarization $\tilde{\mu}$, spread $\tilde{\sigma}$, and the parameter set $\Theta = \{A_0, c, B_T, B_H, a_0, a_1, \dots, a_n\}$ to be estimated by assimilating the observed data and the theoretical model:
\begin{equation}
L(\tilde{\mu}, \tilde{\sigma}, \Theta) = \frac{\lambda_{\mathrm{data}}^{\mu}}{C_{\mathrm{data}}^{\mu}} \mathrm{MSE}_{\mathrm{data}}^{\mu} + \frac{\lambda_{\mathrm{data}}^{\sigma}}{C_{\mathrm{data}}^{\sigma}} \mathrm{MSE}_{\mathrm{data}}^{\sigma} + \frac{\lambda_{\mathrm{eq}}^{\mu}}{C_{\mathrm{eq}}^{\mu}} \mathrm{MSE}_{\mathrm{eq}}^{\mu} + \frac{\lambda_{\mathrm{eq}}^{\sigma}}{C_{\mathrm{eq}}^{\sigma}} \mathrm{MSE}_{\mathrm{eq}}^{\sigma},
\label{eq:loss}
\end{equation}
where $\lambda_{\mathrm{data}}^{\mu}$, $\lambda_{\mathrm{data}}^{\sigma}$, $\lambda_{\mathrm{eq}}^{\mu}$, and $\lambda_{\mathrm{eq}}^{\sigma}$ are weights that balance the relative importance of the four mean squared error terms.

The unknowns $\tilde{\mu}(t)$ and $\tilde{\sigma}(t)$ appear in two places. The first place is in $\mathrm{MSE}_{\text{data}}^{\mu}$ and $\mathrm{MSE}_{\text{data}}^{\sigma}$ of Eq.~\eqref{eq:mse_a}-\eqref{eq:mse_b} and Eq.~\eqref{eq:loss}, where $\tilde{\mu}(t)$ and $\tilde{\sigma}(t)$ are fitted to the known raw observed data $\mu^{*}(t)$ and $\sigma^{*}(t)$. The second place is in $\mathrm{MSE}_{\text{eq}}^{\mu}$ and $\mathrm{MSE}_{\text{eq}}^{\sigma}$ of Eq.~\eqref{eq:mse_c}\eqref{eq:mse_d} and Eq.~\eqref{eq:loss}, where $\tilde{\mu}(t)$, $\tilde{\sigma}(t)$, and $\Theta$ are fitted for the satisfaction of Eqs.~\eqref{eq:mu_discrete} and \eqref{eq:sigma_discrete}. These four mean squared error terms constitute the total loss in Eq.~\eqref{eq:loss}. By minimizing this total loss, all four loss terms are minimized simultaneously. Consequently, the unknowns $\tilde{\mu}$, $\tilde{\sigma}$, and $\Theta$ are then obtained jointly when this total loss reaches its minimum.

The choice of the loss function defined in Eq.~\eqref{eq:loss} is motivated by the following considerations. Firstly, political ideology data, such as the DW-NOMINATE scores, contain inherent noise and possible biases. The computed polarization and spread data, $\mu^{*}(t)$ and $\sigma^{*}(t)$, exhibit considerable discrepancies, especially under the symmetric approximation adopted, as indicated in Fig.~\ref{fig1}. By allowing a balance between data losses and equation losses, we can prevent the model from overfitting to the noisy dataset. Secondly, the two governing equations for the time evolutions of polarization and spread have different levels of accuracy. For instance, the governing equation for $\tilde{\sigma}(t)$ is expected to be less accurate than that for $\tilde{\mu}(t)$. Thus, having four separate terms enables us to manually adjust the weights to achieve an optimal assimilation outcome when necessary.

In minimizing the total loss function Eq.~\eqref{eq:loss}, we employ a gradient-descent-based optimization approach to obtain the optimal parameter set $\Theta$. Specifically, we first provide an initial estimation of $\tilde{\mu}$, $\tilde{\sigma}$, and $\Theta$, then calculate the four mean squared errors, compute the gradients of $\tilde{\mu}$, $\tilde{\sigma}$, and $\Theta$, and update them using the Adam optimizer \cite{Kingma2015} at each iteration. This optimization process is repeated until the total loss converges to an acceptable value.

We employ an updating rate of 0.001 and perform 50,000 iterations to minimize the total loss function and ensure convergence. The weights for the four terms in Eq.~\eqref{eq:loss} are set as $\lambda_{\mathrm{data}}^{\mu} = 1$, $\lambda_{\mathrm{data}}^{\sigma} = 1$, $\lambda_{\mathrm{eq}}^{\mu} = 1$, and $\lambda_{\mathrm{eq}}^{\sigma} = 1$. 
The four corresponding relative root mean square errors (REs) during the minimization process, and the converged REs for polarization and spread with degree $n=7$, are defined as follows:
\begin{subequations}\label{eq:RE}
\begin{align}
\mathrm{RE}_{\mathrm{data}}^{\mu} &= \sqrt{ \frac{ \sum_{k=1}^K \left[ \tilde{\mu}(t_k) - \mu^{*}(t_k) \right]^2 }{ \sum_{k=1}^K \left[\mu^{*}(t_k)\right]^2 } }, \\
\mathrm{RE}_{\mathrm{data}}^{\sigma} &= \sqrt{ \frac{ \sum_{k=1}^K \left[ \tilde{\sigma}(t_k) - \sigma^{*}(t_k) \right]^2 }{ \sum_{k=1}^K \left[\sigma^{*}(t_k)\right]^2 } }, \\
\mathrm{RE}_{\mathrm{eq}}^{\mu} &= \sqrt{ \frac{ \sum_{k=1}^{K-1} \left[ \tilde{\mu}(t_{k+1}) - \tilde{\mu}(t_k) - \tilde{\mu}(t_k) \left( 2\bar{A}(t_k) e^{-2\tilde{\mu}(t_k)/B_H} - \alpha(t_k) \Delta t \right) \right]^2 }{ \sum_{k=1}^{K-1} \left[ \mu^{*}(t_{k+1}) - \mu^{*}(t_k) \right]^2 } }, \\
\mathrm{RE}_{\mathrm{eq}}^{\sigma} &= \sqrt{ \frac{ \sum_{k=1}^{K-1} \left[ \tilde{\sigma}(t_{k+1}) - \tilde{\sigma}(t_k) - \tilde{\sigma}(t_k) \left( \bar{A}(t_k)\left( \frac{4\tilde{\sigma}(t_k)}{\sqrt{\pi}B_T} - 1 \right) - \alpha(t_k)\Delta t \right) \right]^2 }{ \sum_{k=1}^{K-1} \left[ \sigma^{*}(t_{k+1}) - \sigma^{*}(t_k) \right]^2 } }.
\end{align}
\end{subequations}

\begin{figure}[htbp]
    \centering
    \includegraphics[width=0.8\textwidth]{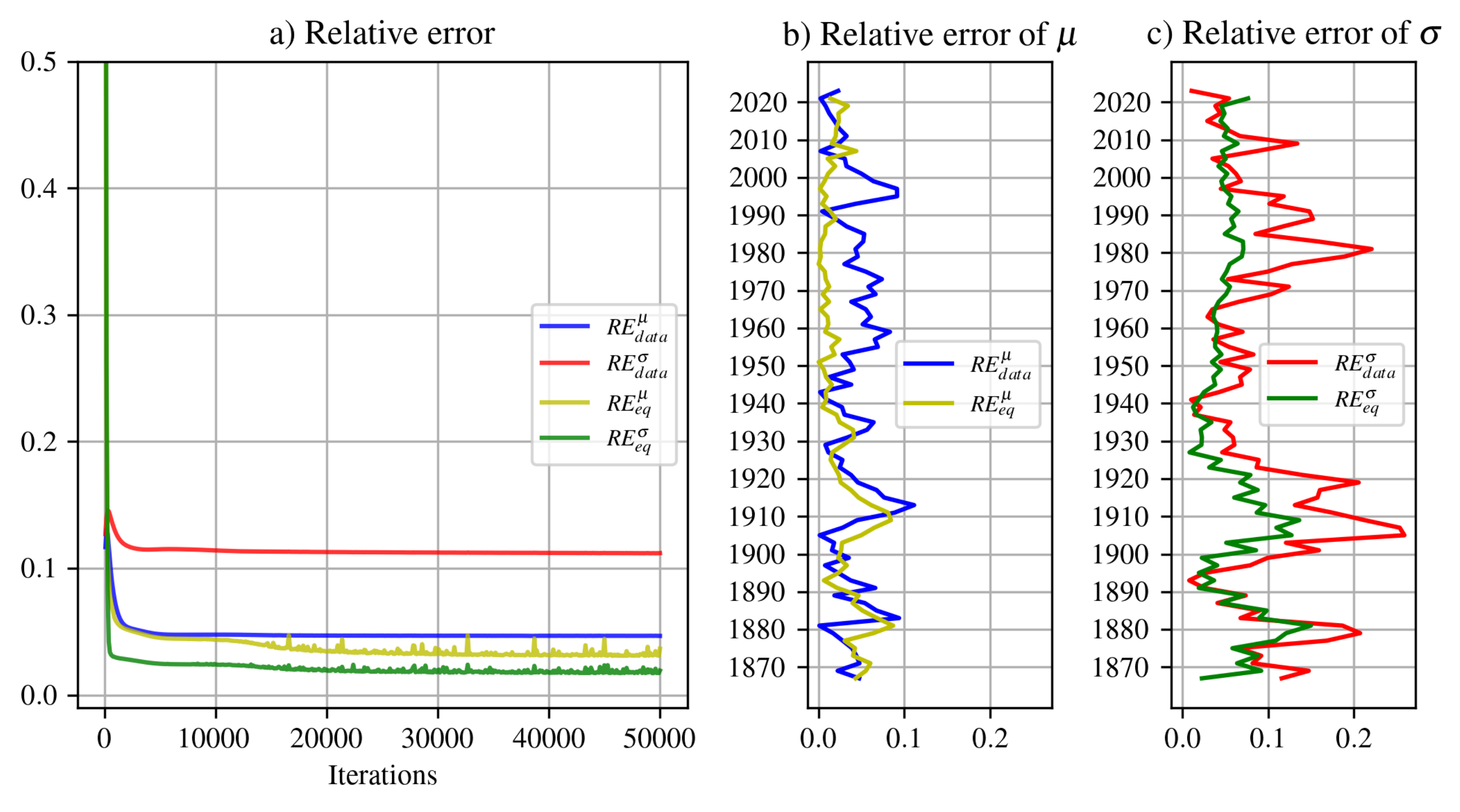}
    \caption{(a) Minimization processes of the four relative root mean squared errors. (b)-(c) Temporal distributions of the relative errors at 50,000 iterations. All results are based on polynomial degree $n=7$.}
    \label{fig2}
\end{figure}

Fig.~\ref{fig2}(a) indicates that the minimization process converges with 50000 iterations and shows that the relative errors for equation satisfactions are smaller than for data discrepancies. Fig.~\ref{fig2}(b) and (c) show that the relative errors for equations, for almost all years, are smaller than those for data discrepancies, for polarization and spread, respectively.

We experiment with four choices of degree $n$, ranging from 5 to 8, for the approximation of $\tilde{\alpha}(x_k)$ with the expansion using Legendre polynomials. During the minimization process, we obtain the polarization $\tilde{\mu}(t)$, the party spread $\tilde{\sigma}(t)$, as well as the parameter set $\Theta=\{A_0,c,B_T,B_H,a_0,a_1,\dots,a_n\}$ as shown in Table~\ref{tab1}. It can be seen that the estimated values of all parameters do not vary significantly for $n=7$ and $n=8$, which indicates that the expansion with $n=7$ is sufficiently accurate for the approximation of $\tilde{\alpha}(x_k)$. 

It should be noted that the seemingly larger number of coefficients $n=7$ is only for the accurate quantification of the strength of the national social norm effect $\tilde{\alpha}(x_k)$, which is an unknown function of time. The political meaning of this obtained function will be provided in the next section inht of the major historical events over the past one and a half centcenturies.
\begin{table}[htbp]
\centering
\caption{Estimated values of parameters for different degrees $n$.}
\label{tab1}
\begin{tabular}{ccccc}
\hline
Parameter & $n=5$ & $n=6$ & $n=7$ & $n=8$ \\
\hline
$A_0$ & 0.551 & 0.535 & 0.591 & 0.603 \\
$c$ & 1.760 & 1.914 & 1.843 & 1.817 \\
$B_H$ & 0.452 & 0.443 & 0.437 & 0.439 \\
$B_T$ & 0.218 & 0.219 & 0.221 & 0.221 \\
\hline
$a_0$ & 0.669 & 0.696 & 0.720 & 0.730 \\
$a_1$ & 0.497 & 0.555 & 0.552 & 0.554 \\
$a_2$ & -0.231 & -0.233 & -0.253 & -0.253 \\
$a_3$ & -0.275 & -0.294 & -0.316 & -0.314 \\
$a_4$ & 0.127 & 0.131 & 0.138 & 0.142 \\
$a_5$ & 0.156 & 0.143 & 0.143 & 0.144 \\
$a_6$ &  & 0.037 & 0.006 & 0.012 \\
$a_7$ &  &  & -0.037 & -0.023 \\
$a_8$ &  &  &  & 0.016 \\
\hline
\end{tabular}
\end{table}

It can be observed that the estimated values of all parameters remain relatively stable between $n=7$ and $n=8$, indicating that the expansion with $n=7$ is sufficiently accurate for approximating $\tilde{\alpha}(x_k)$. It should be noted that the seemingly large number of coefficients ($n=7$) is only required to accurately capture the strength of the national social norm effect, which is an unknown time-dependent function. The political interpretation of the estimated national norm function will be discussed in the next section in light of major historical events over the past one and a half centuries.

\section{Results and Interpretations}

\subsection{Polarization and Spread}

\begin{figure}[htbp]
    \centering
    \includegraphics[width=0.9\textwidth]{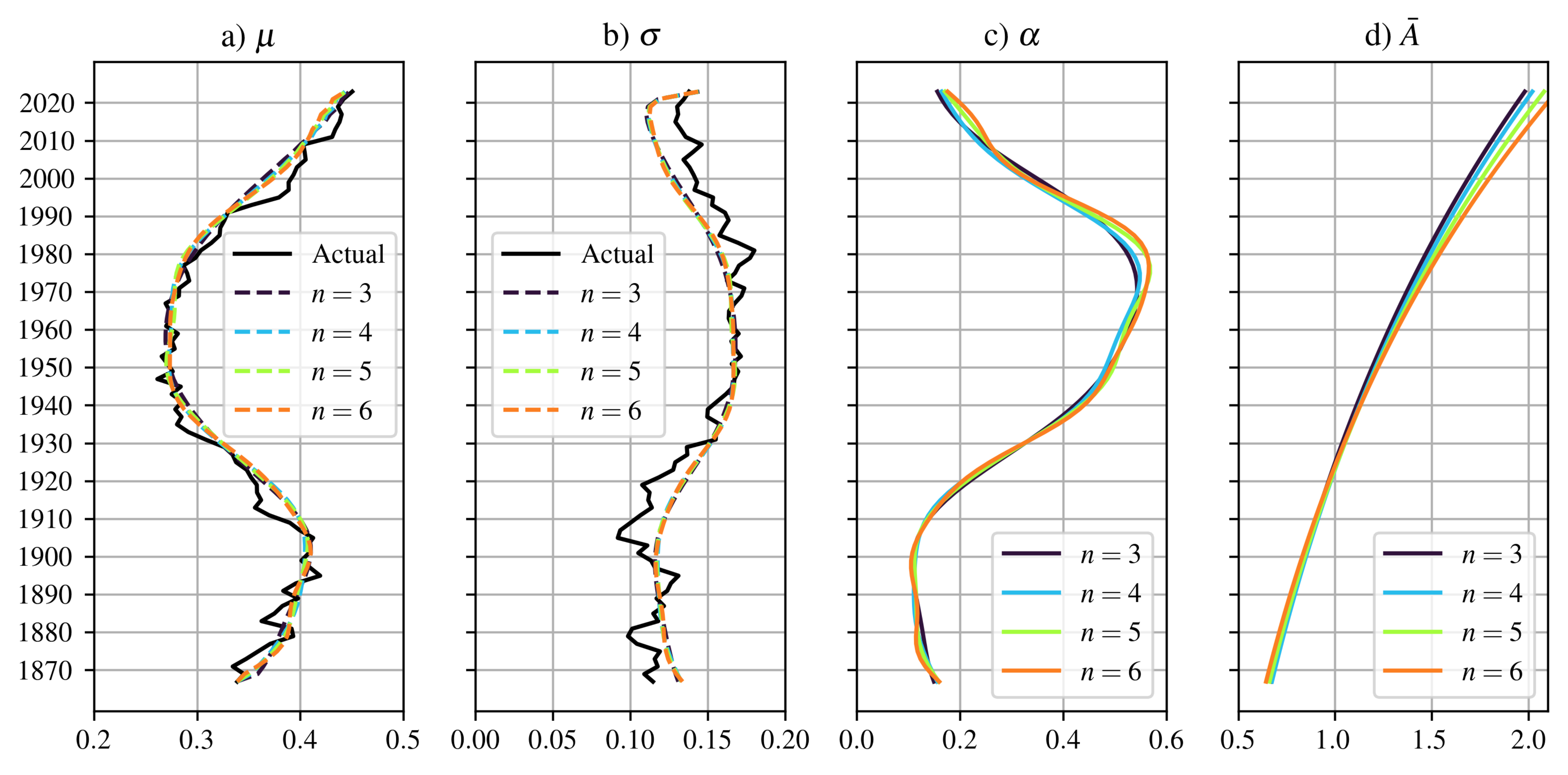}
    \caption{(a) The estimated polarization $\tilde{\mu}(t)$ compared with the corresponding actual data. (b) The estimated spread $\tilde{\sigma}(t)$ compared with the actual data. (c) The estimated strength function of national social norm effect $\tilde{\alpha}(t)$. (d) The estimated influence parameter factor $\bar{A}(t)$. In all plots, $n$ is the number of polynomials used in approximating the strength function of social norm $\tilde{\alpha}(t)$.}
    \label{fig3}
\end{figure}

Fig.~\ref{fig3}(a)(b) compared the estimated polarization $\tilde{\mu}(t)$ and spread $\tilde{\sigma}(t)$ with the observed data. When the polarization increases, the spread decreases, and vice versa, which is consistent with the empirical data and research work of Cole et al.~\cite{Cole2023}. Our model yields a more accurate estimation of polarization than spread. The estimated polarization captures the wavy evolution pattern over the past 154 years, and the estimated spread captures the overall wavy trend as well, with the exceptions of notable deviations from 1900 to 1930 and from 1970 to 2000. With the symmetric approximations adopted in Section 3, we are content with these results at this stage since the objective of this work is only on the overall long timescale wavy patterns of the polarization and spread for the entire recorded historical dataset. The modeling of the asymmetricities will be investigated in the future.

Fig.~\ref{fig3}(c) shows the estimated strength function of the national social norm effect $\tilde{\alpha}(t)$, and Fig.~\ref{fig3}(d) the estimated interaction parameter factor $\bar{A}(t)$. All four plots in Fig.~\ref{fig3} show that the estimated parameters and functions have insignificant differences for the degree of approximation of $\tilde{\alpha}(t)$ using Legendre polynomials from $n=7$ to $n=8$. For simplicity, we will use $n=7$ for the discussions and interpretations below.

\begin{figure}[!h]
    \centering
    \includegraphics[width=0.7\textwidth]{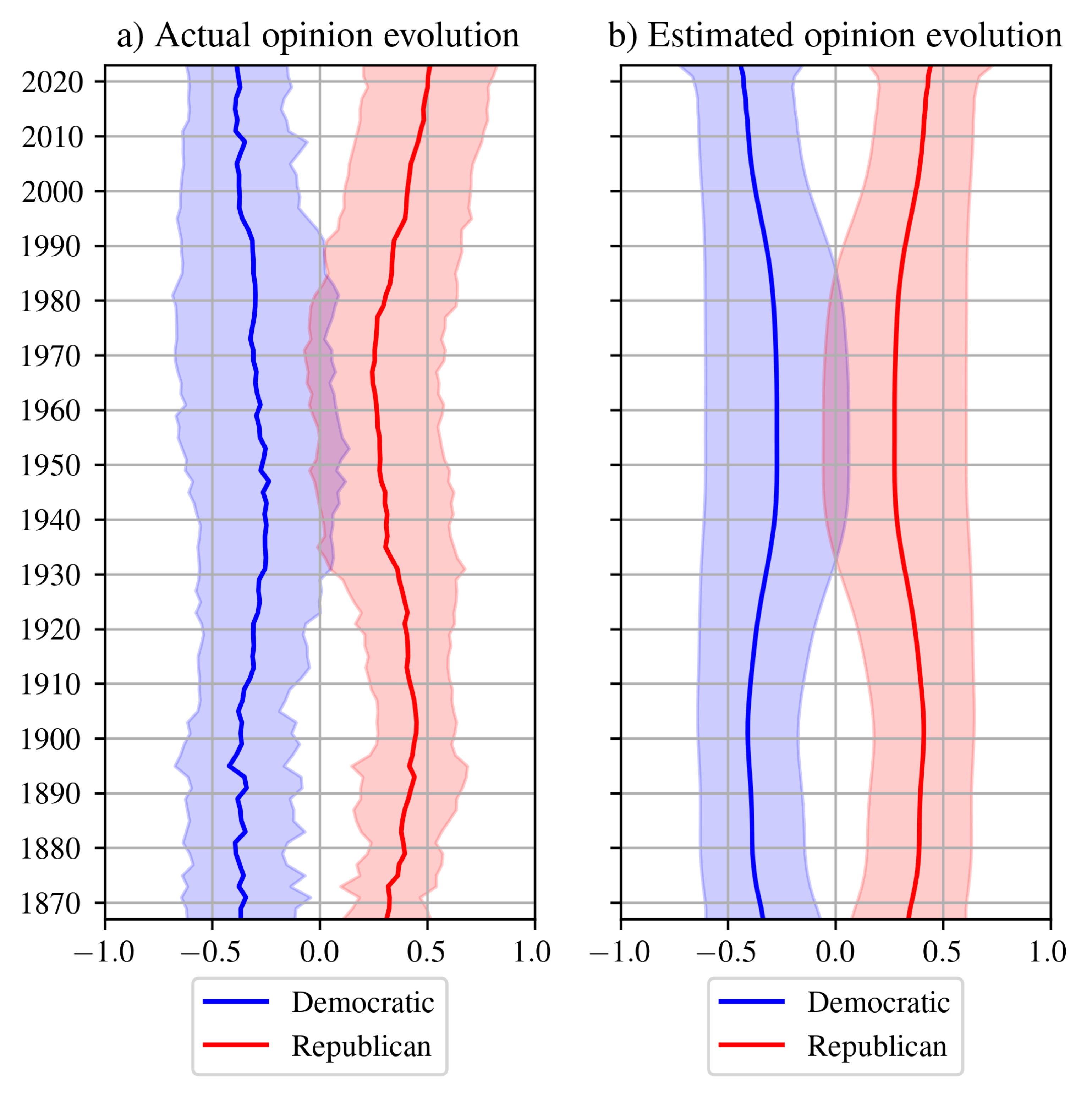}
    \caption{(a) The actual opinion evolution over time. The solid lines represent the party means, and the shaded area represents two standard deviations away from the means. (b) The estimated opinion evolution when $n=7$.}
    \label{fig4}
\end{figure}

The comparison between the evolutions of the observed data and the estimated opinion distributions is shown in Fig.~\ref{fig4}. Typically, stronger polarization corresponds to narrower spread, and vice versa. The slight differences in distributions are primarily due to the symmetry approximation we imposed and the noisy nature of the observed data.

\subsection{Strength of the National Social Norm Effect}

The national social norm strength $\alpha(t)$ can be viewed as the strength of nationwide consensus. Strong consensus occurs when the nation faces a common threat, resulting in ideology clusters attracted to the national norm with a stronger convergence force. Weak consensus occurs during peaceful times when the attractive influence from the national norm is weak, when severe cross-party conflicts emerge, and distant ideology clusters are more likely to diverge.

\begin{figure}[!h]
    \centering
    \includegraphics[width=\textwidth]{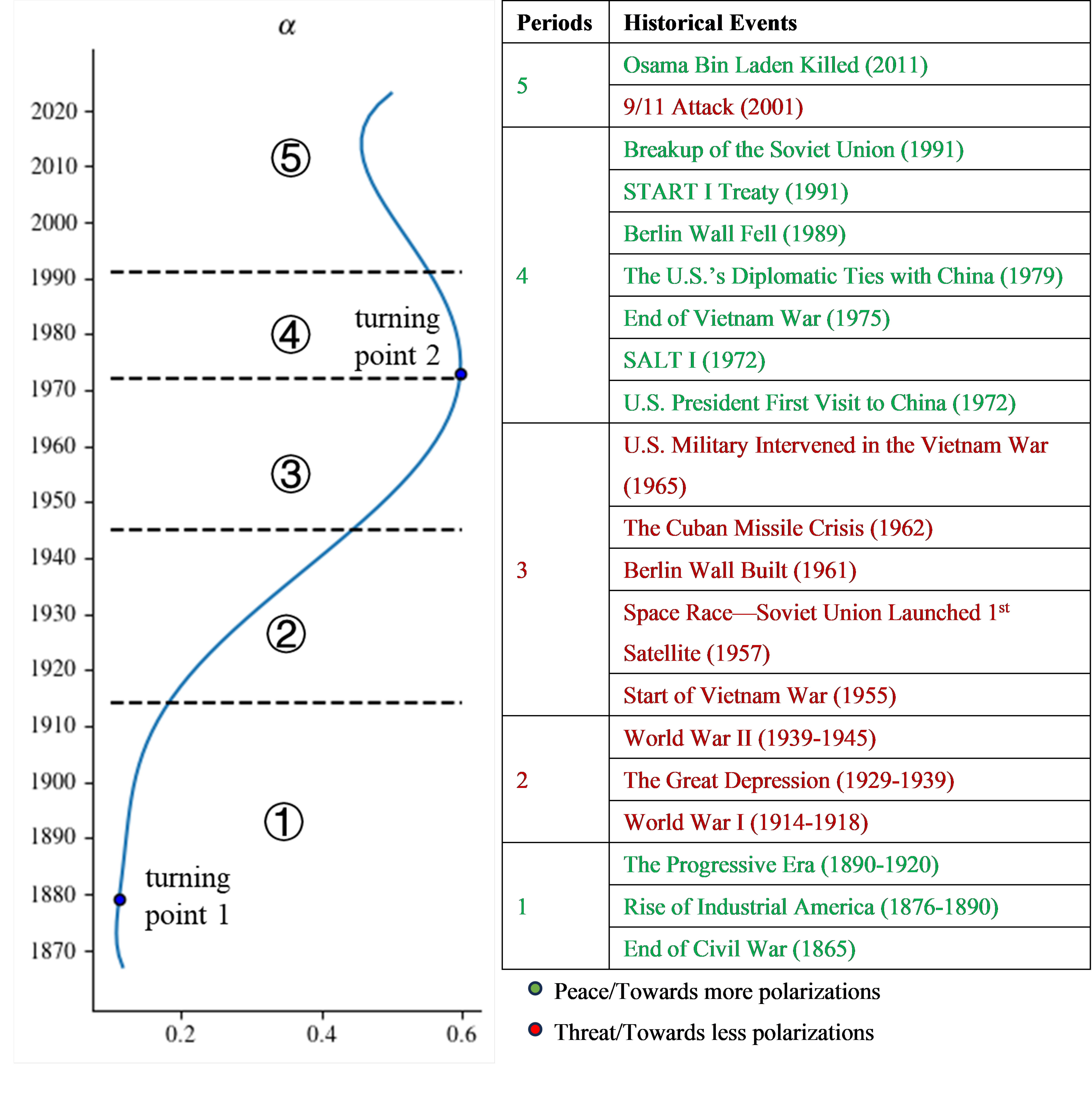}
    \caption{The estimated strength function of the national norm effect $\alpha(t)$ with the division of five periods, alongside the key historical events influencing polarization.}
    \label{fig5}
\end{figure}

In Fig.~\ref{fig5}, we graph the obtained $\alpha(t)$ next to a timeline of notable historical events in the past 154 years. We categorize the events into those that lead to agreements with green color and those that lead to conflicts with red color. A more complete list of events with descriptions is provided in the \textbf{S1 Supporting Information}. Two turning points are presented: the first point around 1880 and the second around 1972. The wavy evolution of the strength of the national social norm is consistent in a historical context by splitting the 154 years into the five periods as shown in Fig.~\ref{fig5}, and the interpretations are provided below.

\textbf{Period 1: Post-Civil War (1868--1914)}  
The Post-Civil War years for the United States were peaceful, as cities were reconstructed and industries began to thrive. While the rise of industrialization produced a class of wealthy industrialists and a prosperous middle class, the working class continued to agonize from unemployment, minimal wages, and pressure from immigrants. Negative sentiment rose as more minor societal conflicts were magnified, resulting in a weak influence from the national norm in the late 19th and early 20th century, as depicted in Fig.~\ref{fig5}. The first turning point of $\alpha(t)$ is around 1880 before the Progressive Movement when a group of activists tried to advocate democracy, expand civil rights, and regulate higher social classes. These political activities slowly strengthened the national norm effect, which agrees with the increasing trend of $\alpha(t)$ from 1890 to 1914 in our result.

\textbf{Period 2: World Wars (1914--1945)}  
The time from 1914 to 1945 witnessed World War I, the Great Depression, and World War II when the Americans were getting united to go through a series of crises. During World War I and II, the national norm strength increased as Americans formed a consensus on defending their country against foreign military forces. On the other hand, studies have shown that during the Great Depression, most Americans were unified and optimistic, mainly due to the enactment of the New Deal. These events together explained the steeply increasing trend of $\alpha(t)$ from 1914 to 1945.

\textbf{Period 3: Early Cold War (1945--1972)}  
In the first thirty years after World War II, a new political consensus was formed concerning the Cold War and anti-communism, causing the national norm strength to continue rising. This consensus peaked around 1972 (the second turning point), indicated by a series of events, including the construction of the Berlin Wall, the Cuban Missile Crisis, the space race, and the Vietnam War, which increased the national norm strength.

\textbf{Period 4: Late Cold War (1972--1992)}  
The once rigid anti-communism consensus began to fragment as the protracted conflict between the United States and the Soviet Union edged towards its conclusion. This period was punctuated by a series of transformative events. In 1972, a notable stride was made when President Nixon signed the SALT I agreement, establishing a framework for limiting strategic armaments. The subsequent year, 1973, witnessed the United States withdraw from the Vietnam War, signaling a retreat from one of the most contentious battlegrounds of the Cold War. The thaw in relations continued as the United States, in 1979, established diplomatic ties with mainland China. The latter years in this period saw a continued diminishment in the confrontations between major superpowers. In 1989, the Berlin Wall fell. In 1991, the United States and Soviet Union signed the START I treaty, agreeing to reduce strategic nuclear arms further. Later that year, the Soviet Union broke up, signaling the denouement of the Cold War. These events show that the conflicts between the United States and communist countries gradually became smaller. With fewer threats outside the country, inner conflicts emerged, and public views began to divide. The slowly decreasing trend of $\alpha(t)$ in this period corresponded to these developments.

\textbf{Period 5: Post-Cold War (1992--2022)}  
The years after the Cold War were relatively peaceful. A few global and national crises still emerged, but the scales of the impacts were much smaller. As a result, we see a decreasing trend in the graph of $\alpha(t)$ from 1991 to 2014. Intriguingly, the resulted $\alpha(t)$ indicates an upward trend in the most recent decade which corresponds to very strong asymmetric opinion distributions of the two parties. Future modeling works are needed to address these asymmetricities.

In summary, a common trend is that national threats are usually corresponding to a growth in the strength of the national norm, and a decrease in polarization and an increase in spread. In contrast, peaceful times are usually corresponding to a decay in the strength, and a rise in polarization and a decrease in spread.

\subsection{Justifications of the Approximations in Section 3}

\textbf{Justification of Approximation 1}:  
The estimated inverse exponential decay parameter $B_H$ is around 0.44, significantly surpassing the range of $\sigma$ from 0.10 to 0.18. Consequently, for most pairs of individuals within the same party, their opinion difference $|B_j - B_i| \sim \sigma < B_H/3$. This approximation is appropriate for capturing the general trend, and the effect of a few extreme cases with short timescale can be neglected.

\textbf{Justification of Approximation 2}:  
The estimated tolerance parameter $B_T$ is around 0.22, and thus, $|B_j - B_i| \sim \sigma < 2B_T$. This suggests that any two individuals in the same party could potentially attract or repel each other based on their pairwise opinion difference. Hence, Approximation 2 is justified.

\textbf{Justification of Approximation 3}:  
For individuals $i$ and $j$ belonging to different parties, $|B_j - B_i| \approx B_R - B_D = 2\mu$, ranging from 0.54 to 0.90, and is larger than $2B_T = 0.44$. While Fig.~\ref{fig1}(a) indicates that a small number of pairs from different parties can have similar opinions that allow attractive interactions, most of the pairs from different parties interact repulsively. Since our model focuses on a qualitative understanding of the general trend of the polarization and spread, it is appropriate to exclude the small portion of individuals whose ideologies lie very close to the national social norm.

\textbf{Justification of Approximation 4}:  
The proportional factor $A(t)$ determines the interaction efficiency in terms of communications among congress members, with parameters $A_0$ and $c$ estimated at 0.6 and 1.8, respectively. We conceptualize interactions among congress members as an integration of cross-district interactions among all citizens. A congress member's opinion level represents a district's average opinion, and an influence exerted from another congress member represents influence from one district to another. Due to rapid advancements in communication technology over the past century and a half, interaction efficiency has exponentially increased, doubling every
\[
K \ln 2 / (c \Delta t) = (78 \times \ln 2) / (1.8 \times 2) \approx 60 \text{ years}.
\]
Although quantitative justification is lacking, this trend aligns with previous qualitative observations~\cite{Cole2023,Gentzkow2016}. The proportional parameter of one individual influencing another in the year of 2022 is
\[
\frac{A_0 e^c}{N \Delta t} = \frac{0.6 \times e^{1.5}}{225 \times 2} \approx 0.006
\]
per year per person for ideology defined in the range from $-1$ to $+1$.

\section{Discussions and Areas for Further Studies}

This work develops a mathematical model aimed at describing the evolution patterns of the opinion distributions within the Democratic and Republican Parties in the U.S. Congress. Grounded in opinion dynamic theory, the model captures the interplay of three time-dependent effects: the national social norm effect, cross-party interactions, and in-party interactions. These effects, whose relative importance varies over time, govern the evolution of polarization and spread within each party. Utilizing an algorithm for theory and data assimilation, we assimilate the model using the U.S. Congressional DW-NOMINATE dataset, which spans over a century and a half of recorded data. The time varying polarization and spread patterns outlined by the model compares well with observations and are consistent with previous studies~\cite{Balietti2021,Cole2023,Ferri2022}.

Notably, while many prior models focus solely on the modeling of either polarization or spread evolution, our model simultaneously captures both of these two interrelated quantities. For example, the ``satisficing'' model proposed by Yang et al.~\cite{Yang2020} models the party polarization with known party spread. Our model models both polarization and spread simultaneously and achieves better estimation on polarization. Moreover, whereas some previous models such as the nonlinear feedback dynamic model~\cite{Leonard2021} compared their results with datasets of limited temporal scope of less than 70 years, our model demonstrates the ability to explain trends across the entire 154 years.

In the model, cross-party interactions augment polarization, while in-party interactions foster spread. The national social norm effect always works to reduce both polarization and spread. By fitting the theory to observational data, we obtain a time-dependent strength function for the national social norm. This function is greater when the nation faces severe threat and smaller when the nation experiences peaceful time. Remarkably, periods of heightened threat correspond to lower polarization and greater spread, whereas periods of peace correspond to higher polarization and reduced spread. This finding aligns well with the important events occurred in the history, at least qualitatively.

It should be noted that future polarization $\tilde{\mu}$ and spread $\tilde{\sigma}$ cannot be determined based on the current dataset. According to Eqs.~\eqref{eq:mu_discrete} and \eqref{eq:sigma_discrete}, the future values of these two variables depend on the future influence factor $A(t)$ and the future strength of the national social norm effect $\alpha(t)$, both of which are currently unknown. If the nation were to encounter a significant threat that leads to an increase in $\alpha(t)$ while $A(t)$ remains relatively constant, both polarization and spread are expected to decrease, and vice versa. Nonetheless, caution is warranted when applying Eqs.~\eqref{eq:mu_discrete} and \eqref{eq:sigma_discrete}, as they are derived based on long-term dynamics and may overlook the impact of short-term effects.

The model maintains internal consistency, with the approximations made in deriving the theory being well-justified. The meanings and values of the four social physical parameters, namely the tolerance parameter, the homophily influence decay parameter, and the two parameters governing the exponentially growing proportional impact factor, are interpretable. From a micro-scale perspective, the model suggests that the repulsion between pairs of individuals when their opinion difference exceeds the tolerance parameter is the root cause of opinion polarization and spread in the political system.

Based on our model, we suggest three potential strategies to mitigate excessive polarization and spread, strengthening shared civic identity to enhance the national norm effect $\alpha(t)$~\cite{VanDerLinden2016}; promoting cross-party engagement to increase the homophily decay parameter $B_H$ and thus lower mutual hostility~\cite{Axelrod2021,Balietti2021}; and encouraging in-party dialogue across differing views to increase the tolerance parameter $B_T$, thereby reducing opinion spread~\cite{Axelrod2021,Balietti2021}.

Our work still contains weaknesses. Significant discrepancies between the model results and the actual data exist, most notably in the asymmetric opinion distribution of the two parties. For example, Fig.~\ref{fig1}(b) and Fig.~\ref{fig4}(a) show a rightward shift of the Republican party compared with the Democratic party in the recent decades.
Modeling such asymmetric patterns may need to remove the symmetric approximation and introduce additional parameters, such as asymmetric tolerance and homophily decay parameters, for both in-party and cross-party dynamics. Studies~\cite{Leonard2021,Rawlings2023,Lelkes2016} indicate that Republican supporters tend to exhibit lower in-party ideological tolerance, while Democratic supporters encompass a more ideologically diverse base. These well-documented insights merit further investigation in subsequent model refinement. However, any future extension---including the addition of asymmetric tolerance and decay parameters---would only be considered if they are supported by robust empirical evidence and lead to significant improvements in the model’s explanatory power.

Considering the nature of this study, we can only limit our conclusions in a qualitative manner before more future work is done. To refine the model and develop operational strategies, future studies especially with empirical experiments are needed. We need to use measurable quantities for the quantification of the tolerance parameter describing the threshold of acceptance to rejection or the reverse, the homophily decay parameter describing the affinity change when people interact with each other, and the strength function of the national social norm effect. For example, the strength of the national social norm effect may be related to critical factors such as economic development, military capability, ideological frameworks, and public health systems.

We invite the collaboration with sociologists and political scientists to improve the model by including more features and offer better interpretations of the relationship between the evolution of opinion distributions and the important social and political events, both past and arising. For example, the mechanisms for the important findings of Jahani et al.~\cite{Jahani2020,Jahani2022} that exposure to common enemies can increase political polarization need to be studied.

With proper extensions, our theory holds promise to study the opinion interactions of three or more groups of people, including communities, parties, countries, etc. By incorporating realistic communication network structures, the theory may be applied to explore dynamics in online opinion spaces. Much work needs to be done, including investigating the influence of opinion elites, the impact of social media, the effects of group norms, and the interplay of multiple interactive opinion topics. Collaboration across disciplines is essential for advancing our understanding of these complex dynamics.

\section*{Supporting Information}
\paragraph{S1 Supporting Information.} 
\textbf{Summary of Notable Historical Events.} 





\clearpage
\appendix
\section*{Appendix A. Derivation of the Equation for Spread}

The spread is defined as the standard deviation of the opinion distribution in each party. Recalling that $B_i<0$ for $i\in D$ and $B_D=-\mu<0$, we have
\begin{align}
\frac{d\sigma^2}{dt} &= \frac{1}{N} \sum_{i\in D} \frac{d(B_i-B_D)}{dt} \cdot 2(B_i-B_D) \notag \\
&= \frac{2}{N} \sum_{i\in D} \frac{dB_i}{dt} \cdot (B_i+\mu) \notag \\
&= \frac{2}{N} \sum_{i\in D} \left( \sum_{j\in D} I_{ji} + NI_{RD} - \alpha B_i \right)(B_i+\mu) \notag \\
&= \frac{2}{N} \sum_{i\in D} \sum_{j\in D} I_{ji}(B_i+\mu) + 2 \sum_{i\in D} I_{RD}(B_i+\mu) - 2\alpha \sigma^2 \notag \\
&= \frac{2}{N} \sum_{i\in D} \sum_{j\in D} I_{ji}(B_i+\mu) - 2\alpha \sigma^2 \notag \\
&= \frac{2}{N} \sum_{i\in D} \sum_{j\in D} I_{ji} B_i - 2\alpha \sigma^2 \notag \\
&= \frac{2A}{N} \sum_{i\in D} \sum_{j\in D} (1 - |B_j-B_i|/B_T)(B_j-B_i)B_i - 2\alpha \sigma^2 \tag{A1}
\end{align}
with the application of the four approximations in Section 3.

By symmetry, the first term in (A1) can also be written as
\[
\frac{2A}{N} \sum_{i\in D} \sum_{j\in D} (1 - |B_j-B_i|/B_T)(B_i-B_j)B_j
\]
Averaging this expression and the first term in (A1), the latter becomes
\[
-\frac{A}{N} \sum_{i\in D} \sum_{j\in D} (1 - |B_j-B_i|/B_T)(B_i-B_j)^2 \tag{A2}
\]

Let $x = B_i$, $y = B_j$, $s = x+y$, $t = x-y$. Since $B_i$ and $B_j$ are independent and normally distributed, the variables $s$ and $t$ are also normally distributed as $s\sim \mathcal{N}(-2\mu, \sqrt{2}\sigma)$ and $t\sim \mathcal{N}(0, \sqrt{2}\sigma)$. 

Denote $p(x) = \frac{1}{\sqrt{2\pi}\sigma} \exp\left( -\frac{(x+\mu)^2}{2\sigma^2} \right)$ as the probability density function for both $x$ and $y$. We can approximate the expression in Eq.~(A2) into an integral:
\begin{align*}
& -\frac{A}{N} \sum_{i\in D} \sum_{j\in D} (1 - |B_j-B_i|/B_T)(B_i-B_j)^2 \\
&= -AN \iint_{-\infty}^{\infty} (1-|x-y|/B_T)(x-y)^2 p(x)p(y) \,dx\,dy \\
&= -AN \iint_{-\infty}^{\infty} (1-|t|/B_T)t^2 \frac{1}{2\pi\sigma^2} e^{-\frac{((s+t)/2+\mu)^2 + ((s-t)/2+\mu)^2}{2\sigma^2}} \frac{1}{2} \,ds\,dt \\
&= -AN \int_{-\infty}^{\infty} \frac{1}{2\sqrt{\pi}\sigma} e^{-\frac{(s+2\mu)^2}{4\sigma^2}} ds \int_{-\infty}^{\infty} (1-|t|/B_T)t^2 \frac{1}{2\sqrt{\pi}\sigma} e^{-\frac{t^2}{4\sigma^2}} dt \\
&= -AN \int_{-\infty}^{\infty} (1-|t|/B_T)t^2 \frac{1}{2\sqrt{\pi}\sigma} e^{-\frac{t^2}{4\sigma^2}} dt \\
&= -\frac{AN}{\sqrt{\pi}\sigma} \int_0^\infty (1-t/B_T)t^2 e^{-\frac{t^2}{4\sigma^2}} dt
\end{align*}

Now it suffices to calculate the integrals $\int_0^\infty t^k e^{-\frac{t^2}{4\sigma^2}} dt$ for $k=2,3$.

Substituting $z = \frac{t^2}{4\sigma^2}$, so that $dt = \sigma dz / \sqrt{z}$, we have:
\begin{align*}
\int_0^\infty t^k e^{-\frac{t^2}{4\sigma^2}} dt &= \int_0^\infty (2\sigma)^k z^{k/2} e^{-z} \frac{\sigma}{\sqrt{z}} dz \\
&= 2^k \sigma^{k+1} \int_0^\infty z^{(k-1)/2} e^{-z} dz \\
&= 2^k \sigma^{k+1} \Gamma\left( \frac{k+1}{2} \right)
\end{align*}
where $\Gamma$ is the gamma function, with $\Gamma(3/2) = \sqrt{\pi}/2$ and $\Gamma(2) = 1$.

Therefore,
\begin{align*}
&-\frac{AN}{\sqrt{\pi}\sigma} \left( \int_0^\infty t^2 e^{-\frac{t^2}{4\sigma^2}} dt - \frac{1}{B_T} \int_0^\infty t^3 e^{-\frac{t^2}{4\sigma^2}} dt \right) \\
&= -\frac{AN}{\sqrt{\pi}\sigma} \left( 4\sigma^3 \frac{\sqrt{\pi}}{2} - \frac{8\sigma^4}{B_T} \right) \\
&= -\frac{2AN\sigma^2}{\sqrt{\pi}} \left( \sqrt{\pi} - \frac{4\sigma}{B_T} \right)
\end{align*}

Substituting into Eq.~(A1), we obtain the governing equation for spread:
\[
\frac{d\sigma}{dt} = \sigma \left[ AN\left( \frac{4\sigma}{\sqrt{\pi} B_T} -1\right) - \alpha \right] \tag{A3}
\]

\paragraph{Acknowledgment} We appreciate the useful assistances and discussions from our colleagues Prof. Miao He and Prof. Wuyue Yang. This work is partially supported by Beijing Institute of Mathematical Sciences and Applications. 

\paragraph{Code and Data Availability} All study data and source code with execution instructions are available on https://doi.org/10.5281/zenodo.15274244.

\paragraph{Author Contributions} X.J.Z. designed research, developed theories and algorithms, analyzed data, and wrote the paper; Y.H. implemented and ran simulation code, derived equations, analyzed data, and wrote the paper; Y.Z. performed research, derived equations, analyzed data, interpreted the results, and wrote the paper. 

\vspace{5mm}
\noindent The authors declare no competing interest.

\end{document}